\documentclass [12pt] {report}
\usepackage{amssymb}
\newcommand {\eqdef} {\stackrel{\rm def}{=}}
\newcommand {\D}[2] {\displaystyle\frac{\partial{#1}}{\partial{#2}}}

\newcommand {\al} {\alpha}

\newcommand {\ga} {\gamma}

\newcommand {\de} {\delta}

\newcommand {\e} {\mathop{\rm e}\nolimits}

\newcommand {\prtl} {\partial}
\newcommand {\fr} {\displaystyle\frac}

\newcommand {\be} {\begin{equation}}
\newcommand {\ee} {\end{equation}}
\newcommand {\ba} {\begin{array}}
\newcommand {\ea} {\end{array}}
\newcommand {\bp} {\begin{picture}}
\newcommand {\ep} {\end{picture}}
\newcommand {\bc} {\begin{center}}
\newcommand {\ec} {\end{center}}
\newcommand {\bt} {\begin{tabular}}
\newcommand {\et} {\end{tabular}}
\newcommand {\lf} {\left}
\newcommand {\rg} {\right}
\newcommand {\iy} {\infty}

\newcommand {\cF} {{\cal F}}

\newcommand {\cR} {{\cal R}}
\newcommand {\cT} {{\cal T}}

\newcommand {\ses} {\medskip}

\newcommand {\bR} {{\bf R}}

\newcommand {\g}  {\stackrel{g\to -g}{\Longleftrightarrow}}

\newcommand {\cE} {{\cal E}}

\newcommand {\bibit} {\bibitem}
\newcommand {\nin} {\noindent}

\def\2#1#2#3{{#1}_{#2}\hspace{0pt}^{#3}}
\def\3#1#2#3#4{{#1}_{#2}\hspace{0pt}^{#3}\hspace{0pt}_{#4}}
\renewcommand{\theequation}{\arabic{sctn}.\arabic{equation}}
\newcounter{sctn}
\def\sec#1.#2\par{\setcounter{sctn}{#1}\setcounter{equation}{0}
                  \noindent{\bf\boldmath#1.#2}\bigskip\par}
\def\Box{{\unitlength1pt\begin{picture}(9,9)
 \put(0,0){\line(1,0){6}}\put(0,0){\line(0,1){6}}
 \put(6,6){\line(-1,0){6}}\put(6,6){\line(0,-1){6}}
 \end{picture}}}
\begin {document}

\begin {titlepage}

\vspace{0.1in}

\begin{center}
{\Large
\bf
Finslerian  Isospin--Nonlinear Equations for Pion
and Spinor Interactions}
\end{center}

\vspace{0.3in}

\begin{center}

\vspace{.15in}
{\large G.S. Asanov\\}
\vspace{.25in}
{\it Division of Theoretical Physics, Moscow State University\\
117234 Moscow, Russia\\

 asanov@newmail.ru

}
\vspace{.05in}

\end{center}

\begin{abstract}

The Finslerian curvature is introduced in the three-dimensional isospin space,
suggesting that the isospin-pion field transforms
according to non-linear Finslerian
invariance
transformations.
The fundamental non-linear realization is associated with the pion field
triplet.
The Finslerian Lagrangian, as well as the implied pion-field selfinteraction,
involves the first but no higher derivatives of the pion field.
While the invariance transformations are now nonlinear,
the conservation laws are still meaningful and sufficient
to get the conservations for the energy-momentum and charges.
The Finsler-invariant pion-nucleon Lagrangian
function is also proposed.
The Finslerian isospin-space symmetry is
a pure interaction symmetry not shared by the
asymptotic fields.
The full Klein-Gordon-Dirac equation is accordingly extended.
Various details of the implied
Finsleroid-structure are formulated in an explicit way.

\ses

\end{abstract}

\end{titlepage}

\vskip 1cm

{\nin\large\bf
Basic Part}
\ses
\ses
\ses
\ses
\ses

\setcounter{sctn}{1}
\setcounter{equation}{0}
{\nin\bf Introduction}
\ses

The Finslerian isospin approach is a representation
of three-dimensional
invariance group in a curved instead of euclidean isospin-space.
Below, we formulate the equations
for possible Finsleroid-mediated symmetry
in the curved isotopic space, based on the
iso-triplet of pions.
So, the triplets are imagined as vectors in the respective
Finsleroid-space.
The
Finsler-geometry ideas enable us to introduce the curvature in the
three-dimensional isospin-space
such that the curvature is constant and positive and is
dependent on a single characteristic parameter.
The resultant Finsleroid-space remains invariant under particular
nonlinear pion-dependent rotations,
henceforth mentioned as the
$\cF_g$--rotations.

We shall start with formulating respective Finslerian field
equations from first principles.
The Finsleroid-type geometry of the curved three-dimensional
isotopic space
replaces now
the ordinary euclidean
geometry of linear pion-represention isotopic space.
We construct the nonlinear Lagrangian such that the Lagrangian is
invariant under
$\cF_g$--rotations.
We get the Finsleroid-mediated
extension of the Klein-Gordon equation,
which involves explicitly a self-interaction of pions
through dependence of the Finslerian metric tensor
components on pion fields.

After that,
we introduce appropriately
the pion-dependence in the nucleon spinor function
$\psi$,
thereby extending the Klein-Gordon-Dirac equations.
The spinor
part is constructed simply by forming isospin-invariant Lagrangian out
of the
$\psi$,
their covariant derivative
$D\psi$,
and an appropriate
pion covariant derivative
$D\pi$.
The function
$\psi$
depends now on both,
the space-time point and the pion field, and transforms under combined
isospin-space transformations.
Remarkably, the right-hand side involves explicitly the
characteristic Finslerian curvature tensor
(the so-called ``Cartan curvature tensor'').
In Appendix,
the initial ingredients of the Finsler geometry are presented.

\ses
\ses

{\nin\bf Formulation of Equations}
\ses
\ses
\ses

Let
$
\pi^q=\pi^q(x)
$
be an isotopic triplet of pions;
the components
$\{
\pi^1(x),
\pi^2(x),
\pi^3(x)
\}
$
are assumed to be real.
Introducing also the notation
$
\pi^q_i=\partial_i\pi^q
=\partial{\pi^q}/\partial{x^i}$
for partial derivatives with respect to
the position point $x^i$,
we construct the Lagrangian
\be
L_{\pi}
=
\fr12r^{ij}
g_{pq}(g;\pi^r)
\pi^p_i\pi^q_j
-\fr12m^2\pi^p\pi_p;
\ee
$m$ denotes the rest mass for the pions;
$
r^{ij}=r_{ij}={\rm diag}(1,-1,-1,-1)
$
is a euclidean metric tensor.
 We may use  the Finslerian metric tensor
$g_{pq}$
(see Appendix below)
and the Finslerian metric function
$K$
(see Appendix below),
so that
\be
\pi^p\pi_p
=K^2(g;\pi^r)
\ee
with
\be
\pi_p
=g_{pq}(g;\pi^r)\pi^q,
\ee
to consider the action integral
\be
S\{\pi^r\}
=\int L_{\pi}d^4x.
\ee
The associated Euler-Lagrange derivatives
\be
\cE_{\pi q}\eqdef\partial_j\D{L_{\pi}}{\pi^q_j}
-\D{L_{\pi}}{\pi^q}
\ee
can explicitly
be calculated from Eqs. (1.1)--(1.5) to yield
$
\cE_{\pi q}=
g_{qp}\cE_{\pi}^p
$
with
\be
\cE_{\pi}^p=
\Box\pi^p
+\3Cqps(g;\pi^r)
\pi^q_i\pi^s_j
r^{ij}
+m^2\pi^p,
\ee
where
\be
\Box=
r^{ij}
\partial_i
\partial_j.
\ee
Since
$\delta S\{\pi^r\}
=0\,\Longrightarrow\,\cE_{\pi}^p=0$,
we arrive at the following
 {\it Finslerian pion field equations}:
\be
\Box\pi^p
+\3Cqps(g;\pi^r)
\pi^q_i\pi^s_j
r^{ij}
+m^2\pi^p
=0,
\ee
where
$C_p{}^r{}_q$ are coefficients given in Appendix below.
The equations are nonlinear and, therefore, introduce
{\it
a Finslerian self-interaction of pions}.
\ses

The associated Hamiltonian is
\be
H\{\pi^r\}=\fr12\int\Bigl[g_{pq}(g;\pi^r)\lf(
\partial_0\pi^p
\partial_0\pi^q
+
\partial_c\pi^p
\partial^c\pi^q\rg)
+m^2
\pi^p\pi_p\Bigr]d^3x
\ee
where
$c=1,2,3$.

Considering
the respective energy-momentum tensor:
\be
T_i{}^j\eqdef
\pi_i^p\D{L_{\pi}}{\pi_j^p}
-\delta_i^jL_{\pi},
\ee
from (1.1) it follows that
\be
T_i{}^j=
g_{pq}(g;\pi^r)
\pi^p_i\pi^{qj}-
\delta_i^jL_{\pi}.
\ee
The conservation law
\be
\partial_jT_i{}^j=0
\ee
\ses
can readily be  verified by direct calculations.
The energy--component reads
\be
T^{00}=
\fr12
g_{pq}(g;\pi^r)
(
\partial^0\pi^p
\partial^0
\pi^q
+
\partial_c\pi^p
\partial^c\pi^q)
+\fr12m^2
\pi^p\pi_p.
\ee

Also, for the currents
\be
j_a{}^n\eqdef
\pi^q\D{L_{\pi}}{\pi_n^p}
t_a{}^p{}_q(g;\pi^r)
\ee
we get the conservation law
\be
\partial_nj_a{}^n=0.
\ee
 {\it The isospin-pion charges}
\be
Q_a=\int{j_a^n} d\Sigma_n=const
\ee
are still conserved,
independently of whether the Finslerian extension is applied,
 or not applied.

Let a field
$\psi$
be an isospinor dublet of four-component space-time spinors,
so that
$\psi
=\{\psi^{\al}\}$.
The Greek indices $\al,\beta...$ label the isospinor
components and take on the values $1,2$; $P,Q=0,1,2,3$.
The notation
${\bf\cT}_P$ will be  used for the ordinary Pauli matrices,
so that
\be
{\bf\cT}_P{\bf\cT}_Q+{\bf\cT}_Q{\bf\cT}_P=2a_{PQ},
\ee
where
$a_{PQ}={\rm diag}(1,1,1)=\de_{PQ}$.
Considering the argument dependence of the spinor to be of the composite
type
\be
\psi=\psi(x,\pi^r(x))
\ee
we shall apply the total derivative
$d_j$:
\be
d_j
\psi^{\beta}(x,\pi^r(x))=
\D
{\psi^{\beta}(x,\pi^r(x))}{x^j}
+
\partial_j\pi^s
\D
{\psi^{\beta}(x,\pi^r(x))}{\pi^s(x)}.
\ee

We introduce,
in agreement with the known methods [1],
{\it
the $\cF$--invariant
isospinor derivative}
\ses
\be
D_j\psi^{\beta}(x,\pi^r(x))=
d_j\psi^{\beta}(x,\pi^r(x))+L^{\beta}{}_{\al j}(g;\pi^r(x))
\psi^{\al}(x,\pi^r(x))
\ee
by the help of
{\it
the $\cF$--invariant
isospinor connection coefficients}
\be
L_j=
L_p(g;\pi^r)
\partial_j\pi^p
\ee
with
\be
L_p(g;\pi^r)
=-\fr18{R^{PQ}}_p(g;\pi^r)
({\bf\cT}_P{\bf\cT}_Q-{\bf\cT}_Q{\bf\cT}_P),
\ee
where
{\it the associated Finslerian
Ricci rotation coefficients}
\be
{R^{PQ}}_p(g;\pi^r)\eqdef
\lf(\prtl_pe_q^Q-\3Cprq e_r^Q\rg)e^{Pq}
\ee
have been used;
$\{e^Q_q\}$
is an appropriate triad.

By the help of the coefficients (1.22)
it is convenient to introduce the isospin-derivative
\be
S_p\psi\eqdef\D{\psi}{\pi^p}+L_p\psi.
\ee
We have
\be
L_p(g;\pi^r)\pi^p\equiv 0, \qquad R^{PQ}{}_p(g;\pi^r)\pi^p\equiv 0.
\ee

The generalized spinor Lagrangian
\be
L_{\psi}
=
-\bar\psi(-i\ga D+M)\psi
\ee
($M$ denotes the rest mass for the fermions)
generates the  extension
\be
(-i\ga D+M)\psi=0
\ee
of the Dirac's equation.
In the component form, the equation (1.26)
reads
\be
-i\ga^jD_j\psi^{\al}+M\psi^{\al}=0.
\ee

Finally,
we construct the objects
\be
E_q\eqdef
e^P_q\cT_P
\ee
and
\be
E^p=g^{pq}E_q
\equiv
e^{Pp}\cT_P.
\ee
For
{\it
the total Lagrangian}
\be
L_{\pi\psi}=
L_{\pi}+L_{\psi}
\ee
we get again the equation (1.28), together with the equation
\be
\Box\pi^p
+\3Cqps(g;\pi^r)
\pi^q_i\pi^s_j
r^{ij}
+m^2\pi^p
=
-\fr14\fr h{1+h}S_q{}^p{}_{ts}(g;\pi^r)
\bar\psi i \ga^n\pi^q_n(E^tE^s-E^sE^t)\psi
\ee
which extends (1.6) by including a due Finslerian
interaction of spinors with pions;
$
S_q{}^p{}_{ts}
$
is the Finsleroid-geometry curvature tensor
(see Appendix below).

\ses
\ses
\ses

{\nin\bf
Discussion}
\ses

Thus we have initiated a systematic development of the Finslerian
nonlinear-invariant
approach
to pion-isospin invariance,
and thereby to pion-pion and nucleon-pion interactions.
It is shown how the Finsler Geometry of the isotopic space
should introduce the interactions.
The approach involves a convenient
nonlinear realization of the isotopic-space rotations.
This realization is associated with the pion field
and is reqarded as a manifestation of a group of
invariance in a curved 3-dimensional isospin space of the Finsleroid type,
leaving invariant the key Finslerian metric function.
Oppositely, we have offered the way of application of
the Finsleroid Metric Function
to processes involving self-interaction of pions.

It can be hoped that this can provide us in future
with new insights into the
pion-nucleon interactions,
as well as into the modern particle interaction models
[2-5].

\ses
\ses
\ses
\ses

\setcounter{sctn}{2}
\setcounter{equation}{0}

{\bf\large APPENDIX.
Finsleroid-Space  $\cE^{PD}_g$  of  Positive-Definite  Type}
\bigskip
\bigskip

Suppose we are given an
$N$--dimensional vector space $V_N$. Denote by $R$ the vectors constituting
the space, so that $R\in V_N$. Any given vector $R$ assigns a particular
direction in $V_N$. Let us fix a member
$R_{(N)}\in V_N$, introduce the
straightline
$e_N$
oriented along the vector
$R_{(N)}$,
 and use this
$e_N$
to serve as a $R^N$--coordinate axis
in $V_N$.
In this way we get the topological product
\be
V_N=          V_{N-1}   \times e_N
\ee
together with the separation
\be
R=\{\bR,R^N\}, \qquad R^N\in e_N \quad {\rm and} \quad \bR\in V_{N-1}.
\ee
For convenience, we shall frequently use the notation
\be
R^N=Z
\ee
and
\be
R=\{\bR,Z\}.
\ee
Also, we introduce a Euclidean metric
\be
q=q(\bR)
\ee
over the $(N-1)$--dimensional vector space
$V_{N-1}$.

With respect to an admissible coordinate basis $\{e_a\}$ in
$V_{N-1}$,
we obtain the coordinate representations
\be
\bR=\{R^a\}=\{R^1,\dots,R^{N-1}\}
\ee
and
\be
R=\{R^p\}=\{R^a,R^N\}\equiv\{R^a,Z\},
\ee
together with
\be
q(\bR)=\sqrt{r_{ab}R^aR^b},
\ee
where $r_{ab}$ are the components of a symmetric positive--definte tensor
defined over $V_{N-1}$.
The indices $(a,b,\dots)$ and
$(p,q,\dots)$ will be specified over the
ranges $(1,\dots,N-1)$ and $(1,\dots,N)$, respectively;
vector indices are up, co--vector indices are down; repeated up--down
indices are automatically summed; the notation $\de^a_b$ will
stand for the Kronecker symbol.
The variables
\be
w^a=R^a/Z, \quad w_a=r_{ab}w^b, \quad w= q/Z,
\ee
where
\be
w\in(-\iy,\iy),
\ee
are convenient whenever $Z\ne0$.
Sometimes we shall mention the associated
metric tensor
\be
r_{pq}=\{r_{NN}=1,~r_{Na}=0,~r_{ab}\}
\ee
meaningful over the whole vector space $V_N$.


Given a parameter $g$ subject to the inequality
\be
-2<g<2,
\ee
we introduce the convenient notation
\be
h=\sqrt{1-\fr14g^2},
\ee
\ses
\be
G=g/h,
\ee
\ses
\be
g_+=\fr12g+h, \qquad g_-=\fr12g-h,
\ee
\medskip
\be
g^+=-\fr12g+h, \qquad g^-=-\fr12g-h,
\ee
so that
\be
 g_++g_-=g, \qquad g_+-g_-=2h,
\ee
\medskip
\be
 g^++g^-=-g, \qquad g^+-g^-=2h,
\ee
\ses
\be
(g_+)^2+(g_-)^2=2,
\ee
\ses
\be
(g^+)^2+(g^-)^2=2,
\ee
and
\be
g_+\g -g_-, \qquad g^+\g -g^-.
\ee

{\it The characteristic
quadratic form}
\be
B(g;R)=Z^2+gqZ+q^2
\equiv\fr12\Bigl[(Z+g_+q)^2+(Z+g_-q)^2\Bigr]>0
\ee
is of the negative discriminant, namely
\be
D_{\{B\}}=-4h^2<0,
\ee
because of Eqs. (2.12) and (2.13).
Whenever $Z\ne0$, it is also convenient to use the quadratic form
\be
Q(g;w)\eqdef B/(Z)^2,
\ee
obtaining
\be
Q(g;w)=1+gw+w^2>0,
\ee
together with the function
\be
E(g;w)
\eqdef
1+\fr12gw.
\ee
The identity
\be
E^2+h^2w^2=Q
\ee
can readily be verified.
In the limit $g\to 0$,
the definition (2.22) degenerates to the
 quadratic form of the input metric tensor (2.11):
\be
B|_{_{g=0}}=r_{pq}R^pR^q.
\ee
Also
\be
Q|_{_{g=0}}=1+w^2.
\ee

In terms of this notation, we propose 
the Finslerian Metric Function
\be
K(g;R)=
\sqrt{B(g;R)}\,J(g;R),
\ee
where
\be
J(g;R)=\e^{\frac12G\Phi(g;R)},
\ee
\medskip
\medskip
\be
\Phi(g;R)=
\fr{\pi}2+\arctan \fr G2-\arctan\Bigl(\fr{q}{hZ}+\fr G2\Bigr),
\qquad {\rm if} \quad Z\ge 0,
\ee
\medskip
\be
\Phi(g;R)=
-\fr{\pi}2+\arctan \fr G2-\arctan\Bigl(\fr{q}{hZ}+\fr G2\Bigr),
\qquad  {\rm if}  \quad Z\le 0,
\ee
\ses\\
or in other convenient forms,
\ses\\
\be
\Phi(g;R)=
\fr{\pi}2+\arctan \fr G2-\arctan\Bigl(\fr{L(g;R)}{hZ}\Bigr),
\qquad  {\rm if}  \quad Z\ge 0,
\ee
\medskip
\be
\Phi(g;R)=
-\fr{\pi}2+\arctan \fr G2-\arctan\Bigl(\fr{L(g;R)}{hZ}\Bigr),
\qquad  {\rm if}  \quad Z\le 0,
\ee
where
\be
L(g;R)
=q+\fr g2Z,
\ee
and
\be
\Phi(g;R)=
\fr{\pi}2
-\arctan{\fr{hq}{A(g;R)}},
\qquad  {\rm if}  \quad Z\ge 0,
\ee
\medskip
\be
\Phi(g;R)=
-\fr{\pi}2
-\arctan{\fr{hq}{A(g;R)}},
\qquad  {\rm if}  \quad Z\le 0,
\ee
\ses\\
where
\be
A(g;R)=Z+\fr12gq.
\ee
\ses\\
This 
 Finslerian Metric Function
has been normalized to show the handy properties
\ses\\
\be
-\fr{\pi}2
\le\Phi\le
\fr{\pi}2,
\ee
\medskip
\medskip
\be
\Phi=
\fr{\pi}2,
\quad {\rm if} \quad q=0 \quad {\rm and} \quad Z>0;
\qquad
\Phi=
-\fr{\pi}2,\quad {\rm if} \quad q=0 \quad {\rm and} \quad Z<0.
\ee

We also have
\ses\\
\be
\cot\Phi=\fr{hq}A, \qquad
\Phi|_{_{Z=0}}=
\arctan \fr G2.
\ee
\ses\\
It is often convenient to use the indicator of sign
$
\epsilon_Z
$
 for the argument $Z$:
\ses
\ses
\be
\epsilon_Z=1, \quad {\rm if}\quad Z>0;
\qquad
\epsilon_Z=-1, \quad {\rm if}\quad Z<0;
\ee

\ses

Under these conditions, we call the considered space {\it the
$\cE^{PD}_g$--space}:
\be
\cE^{PD}_g=\{V_N=V_{N-1}\times e_N;\,R\in V_N;\,K(g;R);\,g\}.
\ee

\ses


The right--hand part of the definition (2.30)
can be considered to be a function
$\breve K$
of  the arguments
$\{g;q,Z\}$,
such that
\be
\breve K(g;q,Z)
=K(g;R).
\ee
We observe that
\be
\breve K(g;q,-Z)\ne \breve K(g;q,Z),\qquad unless \quad g=0.
\ee
Instead, the function $\breve K$ shows the property of
$gZ$--\it parity\rm
\be
\breve K(-g;q,-Z)=\breve K(g;q,Z).
\ee
The $(N-1)$--space reflection invariance
holds true
\be
 K(g;R)\stackrel{R^a\leftrightarrow -R^a}{\Leftrightarrow} K(g;R).
\ee

It is frequently convenient to rewrite the representation (2.30) in the form
\be
K(g;R)=|Z|V(g;w),
\ee
whenever $Z\ne0$, with {\it the generating metric
function}
\be
V(g;w)=\sqrt{Q(g;w)}\,
j(g;w).
\ee
We have
$$
j(g;w)=J(g;1,w).
$$
Using (2.25) and (2.31)--(2.35), we obtain
\be
V'=wV/Q,
\qquad
V''=V/Q^2,
\ee
\ses
\be
(V^2/Q)'=-gV^2/Q^2,  \qquad (V^2/Q^2)'=-2(g+w)V^2/Q^3,
\ee
\ses
\be
j'=-\fr12gj/Q,
\ee
\bigskip\\
and also
\be
\fr12(V^2)'=wV^2/Q,
\qquad\quad
\fr12(V^2)''=(Q-gw)V^2/Q^2,
\ee
\ses
\be
\fr14(V^2)'''=-gV^2/Q^3,
\ee
together with
\be
\Phi'=-h/Q,
\ee
where the prime ($'$) denotes the differentiation with respect to~$w$.

\ses

Also,
\addtocounter{equation}{1}
$$
(A(g;R))^2+h^2q^2=B(g;R)
\eqno(\theequation{a})
$$
and
$$
(L(g;R))^2+h^2Z^2=B(g;R).
\eqno(\theequation{b})
$$
\ses

Sometimes it is convenient to use the function
\be
E(g;w)\eqdef 1+\fr12gw.
\ee


The simple results for these derivatives reduce
the task of computing the components of the associated 
 Finslerian Metric Tensor
to an easy exercise, indeed:
$$
R_p\eqdef\fr12\D{K^2(g;R)}{R^p}:
$$
\ses\ses
\be
R_a=
r_{ab}R^b
\fr{K^2}{B},
\qquad
R_N=(Z+gq)
\fr{K^2}{B};
\ee
\ses
\ses
\ses
\ses
$$
g_{pq}(g;R)
\eqdef\fr12\,
\fr{\prtl^2K^2(g;R)}{\prtl R^p\prtl R^q}
=\fr{\prtl R_p(g;R)}{\prtl R^q}:
$$
\ses
\ses
\ses
\ses
\be
 g_{NN}(g;R)=[(Z+gq)^2+q^2]
\fr{K^2}{B^2},
\qquad
g_{Na}(g;R)=gq
r_{ab}R^b
\fr{K^2}{B^2},
\ee
\ses
\ses
\be
g_{ab}(g;R)=
\fr{K^2}{B}
r_{ab}-g\fr{
r_{ad}R^d
r_{be}R^e
Z
}{q}
\fr{K^2}{B^2}.
\ee
\ses\\
The reciprocal tensor components are
\ses
\be
g^{NN}(g;R)=(Z^2+q^2)
\fr1{K^2},
\qquad
g^{Na}(g;R)=-gqR^a
\fr1{K^2},
\ee
\ses
\be
g^{ab}(g;R)=
\fr{B}{K^2}
r^{ab}+g(Z+gq)\fr{R^aR^b}{q}
\fr1{K^2}.
\ee
\ses\\
The determinant of the 
 Finslerian Metric Tensor
given by Eqs. (2.59)--(2.60) can readily be found in the form
\ses
\be
\det(g_{pq}(g;R))=[J(g;R)]^{2N}\det(r_{ab})
\ee
\ses\\
which shows, on noting (2.31)--(2.33), that
\ses
\be
\det(g_{pq})>0 {\it \quad over~all~the~definition~range} \quad
V_N\setminus 0.
\ee

The associated angular metric tensor
$$
h_{pq}\eqdef g_{pq}-R_pR_q\fr1{K^2}
$$
proves to be given by the components
$$
h_{NN}(g;R)=q^2
\fr{K^2}{B^2},
\qquad h_{Na}(g;R)=-Z
r_{ab}R^b
\fr{K^2}{B^2},
$$
\ses
$$
h_{ab}(g;R)=
\fr{K^2}{B}
r_{ab}-(gZ+q)\fr{
r_{ad}R^d
r_{be}R^e
}q
\fr{K^2}{B^2},
$$
\ses\\
which entails
$$
\det(h_{ab})=\det(g_{pq})\fr1{V^2}.
$$
\bigskip
\bigskip

The  use of the components of the Cartant tensor
(given explicitly in the end of the present section)
 leads,
after rather tedious straightforward calculations, to
the following simple and remarkable result.
\ses

\bf PROPOSITION 1. \it The Cartan tensor associated with the 
 Finslerian Metric Function
\rm(2.30) \it is of the following special
algebraical form:
\be
C_{pqr}=\fr1N\lf(h_{pq}C_r+h_{pr}C_q+h_{qr}C_p-\fr1{C_sC^s}C_pC_qC_r\rg)
\ee
with
\be
C_tC^t=\fr{N^2}{4K^2}g^2.
\ee
\rm
\ses

By the help of (2.65), elucidating the structure of
the curvature tensor
\be
S_{pqrs}\eqdef(C_{tqr}\3Cpts-C_{tqs}\3Cptr)
\ee
results in the simple representation
\be
S_{pqrs}=-\fr{C_tC^t}{N^2}(h_{pr}h_{qs}-h_{ps}h_{qr}).
\ee
Inserting here (2.66), we are led to
\ses

\bf PROPOSITION 2. \it The curvature tensor of the space
$\cE^{PD}_g$ is of the special type
\be
S_{pqrs}=S^*(h_{pr}h_{qs}-h_{ps}h_{qr})/K^2
\ee
with \rm
\be
S^*=-\fr14g^2.
\ee
\ses

{\bf DEFINITION}.\, 
The
 Finslerian Metric Function
 (2.30) introduces an $(N-1)$--dimensional
indicatrix hypersurface  according to the equation
\be
K(g;R)=1.
\ee
We call this particular hypersurface \it the Finsleroid\rm,
to be denoted as
$
\cF^{PD}_g.
$
\ses


Recalling the known formula
$
\cR=1+S^*
$
for the indicatrix curvature,
from (2.70) we conclude that
\be
\cR_{Finsleroid}=h^2=1-\fr14g^2, \qquad\quad 0 < \cR_{Finsleroid} \le 1.
\ee
 Geometrically, the fact that the quantity ~(2.70)
is independent of  vectors~$R$ means that the
indicatrix curvature is  constant. Therefore, we have arrived at
\ses

\bf PROPOSITION 3. \it The Finsleroid
$
\cF^{PD}_g
$
is a constant-curvature space with the
positive curvature value~\rm(2.72).
\ses

Also, on comparing between the result (2.72) and  Eqs. (2.22)--(2.23), we obtain
\ses

\bf PROPOSITION 4. \it The Finsleroid curvature  relates to
the discriminant of the
input characteristic quadratic form~\rm (2.22)
\it
simply as
\be
\cR_{Finsleroid}=-\fr14D_{\{B\}}.
\ee
\rm


Last, we write down the explicit components of the relevant Cartan tensor
\ses\\
$$
C_{pqr}\eqdef \fr12\D{g_{pq}}{R^r}:
$$
\ses
\ses
$$
R^NC_{NNN}=gw^3V^2Q^{-3}, \quad\quad R^NC_{aNN}=-gww_aV^2Q^{-3},
$$
\ses
$$
R^NC_{abN}=\fr12gwV^2Q^{-2}r_{ab}+\fr12g(1-gw-w^2)w_aw_bw^{-1}V^2Q^{-3},
$$
\ses
$$
R^NC_{abc}= -\fr12gV^2Q^{-2}w^{-1}(r_{ab}w_c+r_{ac}w_b+r_{bc}w_a)
 +gw_aw_bw_cw^{-3}\lf(\fr12Q+gw+w^2\rg)V^2Q^{-3};
$$
\ses\\
and
\ses\\
$$
R^N\3CNNN=gw^3/Q^2, \quad\quad\quad R^N\3CaNN=-gww_a/Q^2,
$$
\ses
$$
R^N\3CNaN=-gw(1+gw)w^a/Q^2,
$$
\ses
$$
R^N\3CaNb=\fr12gwr_{ab}/Q+\fr12g(1-gw-w^2)w_aw_b/wQ^2,
$$
\ses
$$
R^N\3CNab=\fr12gw\de_b^a/Q+\fr12g(1+gw-w^2)w^aw_b/wQ^2,
$$
\ses
$$
R^N\3Cabc=  -\fr12g
\lf(\de_a^bw_c+\de_c^bw_a+(1+gw)r_{ac}w^b
\rg)
/wQ
 +\fr12g(gwQ+Q+2w^2)w_aw^bw_c/w^3Q^2.
$$
\ses\\
The components have been calculated by the help of the formulae (2.50)--(2.53).
\ses

The use of the contractions
$$
R^N\3Cabcr^{ac}=-g\fr{w^b}w\fr{1+gw}Q\lf(\fr{N-2}2+\fr1Q\rg)
$$
and
\ses\\
$$
R^N\3Cabcw^aw^c=-g\fr{w}{Q^2}(1+gw)w^b
$$
is handy in many calculations.

Also
\ses\\
$$
R^NC_N=\fr N2gwQ^{-1}, \quad\quad R^NC_a=-\fr N2g(w_a/w)Q^{-1},
$$
\ses
$$
R^NC^N=\fr N2gw/V^2, \quad\quad R^NC^a=-\fr N2gw^a(1+gw)/wV^2,
$$
\ses
$$
C^N=\fr N2gwR^NK^{-2}, \qquad C^a=-\fr N2gw^a(1+gw)w^{-1}R^NK^{-2},
$$
\ses
\ses
\ses
$$
C_pC^p=\fr{N^2}{4K^2}g^2.
$$

For a better understanding of
 structure of the Finsleroid space, the reader is referred to [6].

\ses
\ses
\ses
\ses
\ses

\def\bibit[#1]#2\par{\rm\noindent\parskip1pt
                     \parbox[t]{.05\textwidth}{\mbox{}\hfill[#1]}\hfill
                     \parbox[t]{.925\textwidth}{\baselineskip11pt#2}\par}

\nin{\bf References}
\bigskip

\bibit[1]
G.S. Asanov: \it Finsler Geometry, Relativity and Gauge
 Theories, \rm D.~Reidel Publ. Comp., Dordrecht 1985.

\bibit[2]
 B.W. Lee: \it
Chiral Dynamics, \rm Gordon and Breach, New
York, 1972.

\bibit[3]
 P. Ring and P. Schuck: \it
The Nuclear Many-body Problem,
\rm
Springer, 1980.

\bibit[4]
 J.J J. Kokkedee: \it
The Quark Model,
\rm
Benjamin, New
York, 1969.

\bibit[5]
 T.D. Lee: \it
Particle Physics and Introduction to Field
Theory,
\rm
Harwood Academic, New York, 1981.

\bibit[6] G. S. Asanov: Finsleroids reflect future-past asymmetry,
-
\it Rep. Math. Phys. \bf 47 \rm(2001), 323.

\end {document}